\documentclass[usegraphicx,usenatbib]{mn2e}

\usepackage{times}
\usepackage{amsmath}
\usepackage{amssymb}

\begin{document}

\title[Absorption in the Chandra Deep Fields] {Radiation pressure, absorption and AGN feedback in the Chandra Deep Fields}
\author[S.I. Raimundo et al.]
{\parbox[]{7.in}{S.~I. Raimundo$^{1}$\thanks{E-mail: 
sijr@ast.cam.ac.uk}, A.~C. Fabian$^{1}$, F.~E. Bauer$^{2}$, D.~M. Alexander$^{3}$, W.~N. Brandt$^{4}$, B.~Luo$^{4}$, R.~V. Vasudevan$^{4}$ and Y.~Q. Xue$^{4}$\\
\footnotesize
$^{1}$Institute of Astronomy, Madingley Road, Cambridge CB3 0HA\\
$^{2}$Pontificia Universidad Cat\'olica de Chile, Departamento de Astronom\'ia y Astrof\'isica, Casilla 306, Santiago 22, Chile\\
$^{3}$Department of Physics, Durham University, Durham, DH1 3LE\\
$^{4}$Department of Astronomy \& Astrophysics, 525 Davey Lab, The Pennsylvania State University, University Park, PA 16802, USA\\}
}

\maketitle
\begin{abstract}
The presence of absorbing gas around the central engine of Active Galactic Nuclei (AGN) is a common feature of these objects. Recent work has looked at the effect of the dust component of the gas, and how it enhances radiation pressure such that dusty gas can have a lower effective Eddington limit than ionised gas.
In this work, we use multi-wavelength data and X-ray spectra from the 2 Ms exposures of the \textit{Chandra} Deep Field North and \textit{Chandra} Deep Field South surveys, to characterise the AGN in terms of their Eddington ratio ($\lambda$) and hydrogen column density (N$_{\rm H}$). Their distributions are then compared with what is predicted when considering the coupling between dust and gas. Our final sample consists of 234 objects from both fields, the largest and deepest sample of AGN for which this comparison has been made up to date.
We find that most of the AGN in our sample tend to be found at low Eddington ratios (typically $10^{-4}<\lambda<10^{-1}$) and high N$_{\rm H}$ ($>$10$^{22}$cm$^{-2}$), with black hole masses in the range $\sim(10^{8}-10^{9})M_{\odot}$. Their distribution is in agreement with that expected from the enhanced radiation pressure model, avoiding the area where we would predict the presence of outflows.
We also investigate how the balance between AGN radiation pressure and gravitational potential influences the behaviour of clouds in the galactic bulge, and describe a scenario where an enhanced radiation pressure can lead to the fundamental plane of black hole/galaxy scaling relations.
\end{abstract} 
\section{Introduction}
Observations of Active Galactic Nuclei (AGN) have revealed that many of them are partially obscured by material in our line of sight, within the inner tens of parsecs of the central engine (e.g. \citealt*{risaliti99} and \citealt*{maiolino&risaliti07} for a review).
This obscuring material will influence both the AGN and our observations, and the study of its properties is fundamental for an unbiased understanding of AGN physics.
The gas around the nucleus will be under the effect of the inward gravitational force of the supermassive black hole and the outward pressure of the radiation emitted in the central region. By investigating the balance between these two forces, one can predict the behaviour of the gas. The Eddington luminosity, $L_{\rm E}$, is defined as the value at which the radiation pressure balances the gravitational force of the black hole: $L = L_{\rm E}$, with $L_{\rm E}=4\pi Gm_{\rm p}cM_{\rm BH}/\sigma_{\rm T}$, where G is the gravitational constant, m$_{\rm p}$ the proton mass, c the speed of light, $M_{\rm BH}$ the mass of the black hole and $\sigma_{\rm T}$ the cross-section for Thomson scattering. The Eddington ratio ($\lambda = L/L_{\rm E}$), is then a measure of the balance between these two forces, for a certain $M_{\rm BH}$.
In the presence of dust, the gas couples with the dust grains via Coulomb interactions and the cross-section for the interaction with photons is considerable enhanced. The effective cross-section for dusty gas can be defined as $\sigma_{\rm d}=A\sigma_{\rm T}$ with a boost factor $A$ (\citealt{fabiancelottierlund06}, \citealt*{fabian&vasudevan08}, \citealt{fabian09}).
We can then determine an effective Eddington ratio for dusty gas as a function of the classical Eddington ratio: 
\begin{eqnarray*} \label{edd_eff}
\lambda_{\rm eff} = \frac{L\sigma_{\rm d}}{4\pi Gm_{\rm p}cM_{\rm BH}} = A\lambda.
\end{eqnarray*}
It follows that, $\lambda = 1/A$ is now the limit at which the radiation pressure from the black hole is able to expel the mass of dusty gas around it. 
\cite{fabian&vasudevan08} and \cite{fabian09} explored the effective Eddington limit for dusty gas by investigating the properties of AGN samples in the \textit{Chandra} Deep Field South (CDF-S), Lockman Hole and in the local Universe, and found that the objects tend to lie below their effective Eddington limit, as expected.

AGN are expected to interact with the surrounding galaxy, affecting its evolution. In fact, there is evidence of a close interaction between galaxies and the supermassive black holes at their centres. Various studies have found correlations between the mass of the black hole and several properties of the galaxy when looking at samples in the local Universe. \citet{kormendy&richstone95} summarised the evidence for the presence of massive dark objects in the centres of galaxies and established a dependence between the masses of those objects and the galaxy luminosity/mass, a conclusion later strengthened by \cite{magorrian98} when looking at a larger sample of galaxies. \cite{gebhardt00} and \cite{ferrarese&merritt00} found that the black hole mass correlates with the velocity dispersion, and later on, correlations with other properties were determined (e.g.: S\'ersic index - \citealt{graham01}). It has been suggested that these correlations are projections of a more fundamental relation or ``fundamental plane''. This argument is based on what is known for elliptical galaxies, where a fundamental plane was determined, involving the luminosity/effective radius, the velocity dispersion and the surface brightness (\citealt{dressler87,djorgovski&davis87}). \cite{marconi&hunt03}, found evidence of a correlation between the residuals in the $M_{\rm BH}-\sigma$ relation and the galaxy bulge effective radii. More recently, \cite{hopkins07} studied in detail the dependence of $M_{\rm BH}$ on the various galactic properties, establishing a fundamental plane of the form $M_{\rm BH}\propto \sigma^{3.0\pm 0.3}R^{0.43\pm 0.19}$ or $M_{\rm BH}\propto M^{0.54\pm 0.17}_{*}\sigma^{2.2\pm 0.5}$, reinforcing the idea of a very strong connection between the black hole and the evolution of the galaxy.

In this work we investigate how the radiation emitted by the AGN interacts with the obscuring gas around the central black hole, and determine how this affects the structure of the host galaxy. We start by determining the obscuration properties of a large sample of AGN detected in the deepest X-ray surveys of the \textit{Chandra} Deep Field North (CDF-N) \citep{alexander03} and \textit{Chandra} Deep Field South (CDF-S) \citep{luo08}. The properties of our objects are then compared with predictions from the enhanced radiation pressure model, as in \cite{fabian&vasudevan08} and \cite{fabian09}, but with a deeper exposure, to enable the analysis of a larger number of AGN and a better signal to noise ratio in the X-ray spectral analysis. Finally, we investigate the role of the radiation/dusty gas interaction in the properties of the host galaxy, namely, on establishing the scaling relations between the black hole and the galaxy.
We adopt standard cosmological parameters of H$_0 = 70$ km s$^{-1}$ Mpc$^{-1}$, $\Omega_{\rm m} = 0.27$ and $\Omega_{\rm \Lambda} = 0.73$.
\begin{figure}
\centering
\includegraphics[width=0.8\columnwidth]{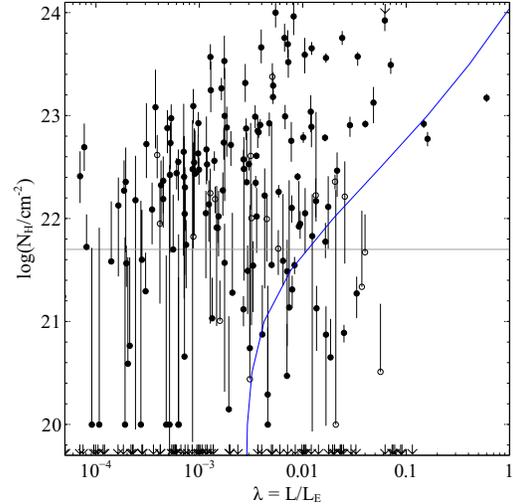}
\caption {Hydrogen column density (N$_{\rm H}$) versus Eddington ratio ($\lambda$), for AGN with $z<1$. The filled circles represent objects with measured spectroscopic redshifts and the open circles objects with photometric redshifts. The downward arrows point to the Eddington ratios of objects with negligible absorption (log N$_{\rm H} \sim 19.0$). The blue line represents the effective Eddington limit when accounting for the presence of dust, and the horizontal grey line is an approximate limit for the column density of a dust lane ($N_{\rm H}\sim 5\times 10^{21}$cm$^{-2}$). Outflows or transient absorption are expected to happen above the limit for dust lanes and to the right of the blue line. Objects tend to lie in the upper left region of the plot, with high column densities, avoiding the outflow area.
The vertical error bars represent the 1$\sigma$ deviations in N$_{\rm H}$. Errors in $\lambda$ are of the order of 30\%.}
\label{NhEddlowz}
\end{figure}
\section{Data Analysis}
To compare the data with the dusty gas model, we used the point sources found with the 2\thinspace Ms exposure in the CDF-N \citep{alexander03} and the 2\thinspace Ms exposure in the CDF-S \citep{luo08}. 
Spectroscopic redshifts for the CDF-S were collected by \cite{luo10} and by \cite{rafferty10} for the CDF-N. Photometric redshifts for the CDF-S are also from \cite{luo10}, with a typical accuracy of $|\Delta z|/(1+z)\approx 7\%$, and from \cite{xue10} for the CDF-N, with an accuracy of $\approx 11\%$.
X-ray spectra for our sample were extracted from the CDF-N and CDF-S datasets using the IDL routine ACIS {\it Extract} (\citealt{broos10}), which incorporates several advancements over standard CIAO procedures. The spectra were fitted in {\sc xspec} (v12.2.1o; \citealt{arnaud96}) with an absorbed powerlaw model {\tt tbabs}({\tt ztbabs}+{\tt pegpwrlw)}, allowing the photon index to change and with the normalisation determined by the 0.5-8 keV observed flux. The {\tt pegpwrlw} model was used to allow an unbiased determination of the normalisation. A full account of the spectral analyses of the CDF sources will be presented in Bauer et al. 2010 (in prep).
Objects for which the best-fitted parameters tended to extreme non-physical values were excluded, (for example, when the two fitting parameters, $\Gamma$ and $N_{\rm H}$ tend to the limits imposed in the fit), since this is an indication that the model adopted does not provide an accurate description of the data.
The X-ray intrinsic (absorption corrected) luminosity in the 2-10 keV band for each source, was obtained from the spectral fit parameters and used to identify AGN: we select objects with $L_{\rm X}>10^{41}$erg\thinspace s$^{-1}$, since the X-ray emission can be contaminated by star formation in lower luminosity sources. The value of hydrogen column density ($N_{\rm H}$) was obtained from the X-ray spectral fitting with the models described above.

To obtain the Eddington ratio for the AGN, the bolometric luminosity and the mass of the black hole are needed. The X-ray luminosities in the 2-10 keV band were converted to bolometric luminosities using a bolometric correction of 19 for objects with an X-ray photon index, $\Gamma$, lower than 1.9 and a correction of 55 otherwise, as done in \cite{fabian09}. This correction is based on \cite{winter08}, where a correlation is found between $\Gamma$ and Eddington ratio in various observations of individual objects. The total absolute rest-frame $K$-band magnitude of the galaxy was obtained by \cite{xue10}  
from SED fitting. This value was corrected for the nuclear emission from the AGN using an X-ray luminosity and $N_{\rm H}$ dependent scaling factor (Eq.~1 from \citealt{vasudevan09}), and we consider that the galaxy host $K$-band light is dominated by the bulge, following studies which found AGN hosts to be typically early type galaxies \citep{kauffmann03} with bulge dominated morphologies \citep{grogin05}. This may however be a simplification, a recent study showed that X-ray selected AGN hosts show a broad range of morphologies, with an intermediate distribution between bulge and disc dominated galaxies \citep{gabor09}. The masses were obtained using the scaling relation between the black hole mass and the bulge absolute $K$-band magnitude: log$($M$_{\rm BH}/$M$_{\odot})=-0.37(\pm 0.04)($M$_{\rm K} + 24) + 8.29(\pm 0.08)$ \citep{graham07}. The possible evolution of the relation between black hole mass and $K$-band magnitude will be discussed in section 3.1.

The objects in our sample were selected from both {\it Chandra} Fields, and have redshifts $z<1$. There are 253 AGN with measured $K$-band magnitudes and redshifts below one. Out of these, 19 are excluded due to problems in the fit (fitting parameters converging  to extreme values). In summary we have a sample of 234 AGN, out of which 207 have spectroscopic redshifts and 27 have photometric redshifts.
\begin{figure}
\centering
\includegraphics[width=0.8\columnwidth]{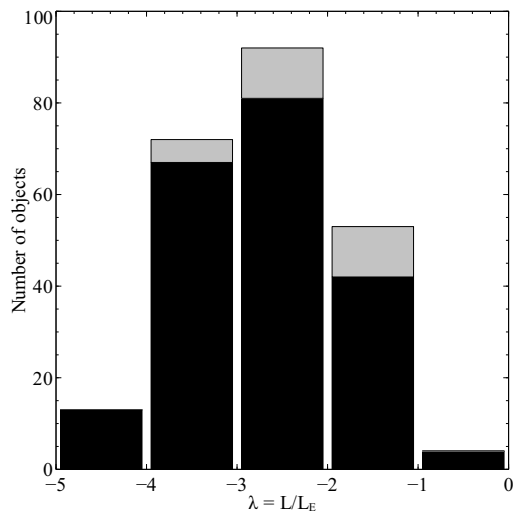}
\centering
\caption {Number of objects with $z<1$ binned in terms of their Eddington ratios ($\lambda$). Black histogram represents sources with spectroscopic redshifts, grey histogram adds the sources with photometric redshifts only. AGN have a wide range of Eddington ratios, varying mainly between $10^{-4}-10^{-1}$, as also found by \citet{babic07}.}
\label{EddBins}
\end{figure}
\section{Discussion}
We plot our results in Fig.~\ref{NhEddlowz} in the same way as in \cite{fabian&vasudevan08} and \cite{fabian09}. The blue line represents the effective Eddington limit for dusty gas, obtained using {\sc cloudy} (details in \citealt{fabiancelottierlund06}, \citealt{fabian&vasudevan08}). We expect the objects to the right of the line to have outflows or to have the absorbing gas further away from the central black hole, so that they are under the effect of a larger gravitational mass. If they are associated with dust lanes, (e.g. \citealt{matt00}), the $N_{\rm H}$ value will not be intrinsic AGN absorption only, it will be caused by gas further out in the galaxy as well. If there is a dust lane, its mass of gas has a maximum limit, so the column density cannot be larger than a certain value. We use as reference the value of $N_{\rm H} = 5\times 10^{21}$cm$^{-2}$ as in \cite{fabian&vasudevan08} and plot it in Fig.~\ref{NhEddlowz} as a grey horizontal line. To the right of the blue line and above the grey line, we expect to have objects experiencing outflows or variable absorption. The Compton-thick objects ($N_{\rm H} \gtrsim 1.5\times 10^{24}$cm$^{-2}$) never see the source above the effective Eddington limit, so their absorption is long-lived for $\lambda < 1$.
Most of our objects lie in the upper left region of the plot, with a high column density (log N$_{\rm H}\gtrsim 22$) and low Eddington ratio ($\lambda \lesssim 10^{-1}$). More than 95$\%$ of the sources with log N$_{\rm H}>5\times10^{21}$cm$^{-2}$  avoid the area where we would expect to see outflows (to the right of the blue line and above the horizontal grey line), which means that the absorbing clouds around them are long-lived.
Considering the error bars, there are only two objects with an Eddington ratio clearly above the effective Eddington limit. This is unlikely to be a selection effect, since for a given $M_{\rm BH}$ the sources with higher Eddington ratios should be in general easier to detect. 
\begin{figure}
\centering
\includegraphics[width=0.8\columnwidth]{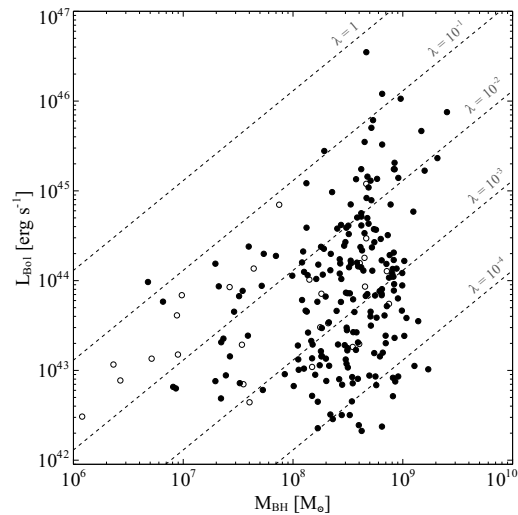}
\centering
\caption {Bolometric luminosity as a function of mass for every AGN with $z<1$. The filled circles represent AGN with measured spectroscopic redshifts and the open circles AGN with photometric redshifts only. The dashed lines separate between different Eddington ratios. Most of the AGN have $M_{\rm BH}\sim (10^{8}-10^{9})M_{\odot}$.}
\label{EddLbol}
\end{figure}

We explored the effects of using different bolometric corrections. Both a constant bolometric correction of $\sim 30$, or an Eddington ratio dependent bolometric correction \citep{vasudevan&fabian07} reproduce the same general properties as deduced from Fig.~\ref{NhEddlowz}. The position of the points change relatively, but there are only two clear points with $N_{\rm H}>5\times10^{21}$cm$^{-2}$ above the effective Eddington limit and those are the same objects as in Fig.~\ref{NhEddlowz}.  
The Eddington ratio distribution of our sample ($z<1$) can be seen in Fig.~\ref{EddBins}. It peaks at around ($10^{-3}-10^{-2}$) with most of the sources having Eddington ratios between ($10^{-4}-10^{-1}$) in agreement with what was found by \cite{babic07} when looking at AGN in the CDF-S.
In summary, our sample can be characterised in terms of its overall properties in Fig.~\ref{EddLbol}; the black hole mass for the majority of the sources is in the range $M_{\rm BH}\sim(10^{8}-10^{9})M_{\odot}$, with bolometric luminosity $L_{\rm bol}\sim (10^{42}-10^{45})$erg\thinspace s$^{-1}$ and relatively low Eddington ratio. We obtained black hole masses that are in general higher than the ones found in \cite{ballo07}, the majority of their sources have $M_{\rm BH}\sim(10^{7}-10^{8.5})M_{\odot}$. This discrepancy can be partially explained by the different scaling relation adopted in their work: the R-band magnitude relation from \cite{mclure&dunlop02}, which yields a conservative lower limit for the black hole masses.
\subsection{Comparison with other samples}
\begin{figure}
\centering
\includegraphics[width=0.8\columnwidth]{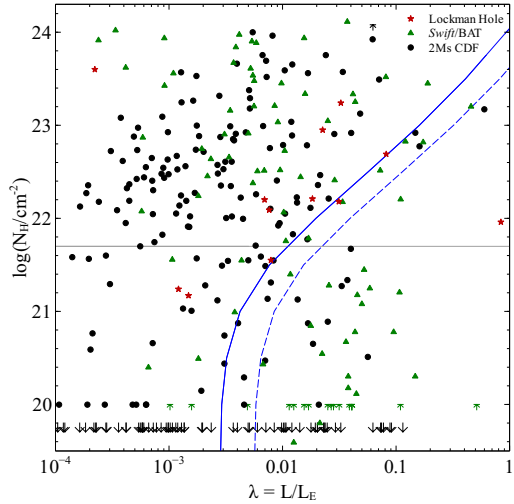}
\caption {Hydrogen column density (N$_{\rm H}$) versus Eddington ratio ($\lambda$) summary plot of results from the Lockman Hole \citep{fabian&vasudevan08}, \textit {Swift}/BAT \citep{fabian09} and 2 Ms \textit {Chandra} Deep Fields (this work). Sources from the 9 month \textit {Swift}/BAT catalogue of local AGN are represented with green triangles, Lockman Hole with red stars, and black circles for the AGN in the \textit {Chandra} Deep Fields. Upward arrows are upper limits for some of the objects in the \textit {Swift}/BAT catalogue. The dashed blue line considers the mass of intervening stars when calculating the effective Eddington limit.}
\label{NhEddall}
\centering
\end{figure}
We compared our results with those previously found by \cite{fabian&vasudevan08} and \cite{fabian09}. These authors looked at local samples of AGN from the \textit{Swift}/BAT survey \citep{winter08}, and higher redshift ($0.5<z<1$) samples from the Lockman Hole (\citealt{mainieri02}; \citealt{mateos05}) and from the 1\thinspace Ms exposure in the CDF-S \citep{tozzi06}.
In Fig.~\ref{NhEddall}, we compile the results from the present work, \cite{fabian&vasudevan08} and \cite{fabian09}. The 1\thinspace Ms CDF-S results are excluded due to our current analysis of a deeper observation in this field (2\thinspace Ms).
For the gas further away from the black hole, (hundreds of parsecs to kpc), the enclosed mass will include stars, and the effective Eddington limit will be higher. We follow \cite{fabian09} and plot a dashed line to indicate a factor of two increase in the enclosed mass. Analysing all the results, with a larger number of sources at different redshifts, we obtain the same conclusions stated previously: the AGN tend to avoid the area where outflows are expected (in the upper right corner of the plot). More than 90$\%$ of the sources with log N$_{\rm H}>5\times10^{21}$cm$^{-2}$ are located to the left of the solid blue line, and more than 98$\%$ lie to the left of the dashed blue line.

\begin{figure}
\begin{center}
\includegraphics[width=0.8\columnwidth]{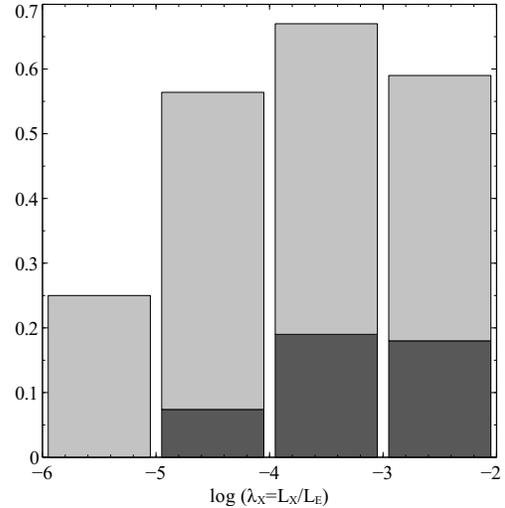}
\caption {Fraction of objects binned as a function of their X-ray Eddington ratio ($\lambda_{\rm X}$). The light grey histogram represents the fraction of sources with log $N_{\rm H} > 22$ and the dark grey sources with log $N_{\rm H} > 23$.}
\label{GammaNhBins}
\end{center}
\end{figure}

We also limited our sample to log $L_{\rm X} > 41.5$, to have approximately the same range of X-ray luminosities as the AGN in the \textit{Swift}/BAT study of \cite{winter09}, and to be able to compare some of our $z < 1$ sample properties with what is found for their local AGN.
The fraction of absorbed sources (log N$_{\rm H}>22$) and the fraction of highly absorbed sources (log N$_{\rm H}>23$) as a function of the X-ray Eddington ratio are plotted in Fig.~\ref{GammaNhBins}. The absorbed AGN constitute, in general, the largest fraction of sources, as found in the local sample of \cite{winter09}. The fraction of absorbed sources at low $\lambda_{\rm X}$ is nevertheless lower than in the local sample, but this is due to the fact that the \emph{Swift}/BAT sample has more obscured AGN at low luminosities than our own \citep{mullaney10}. 

From our study, we predict a decrease in the fraction of obscured objects with increasing Eddington ratio, in particular for the highest values of $\lambda$. This trend would be seen for objects with log $N_{\rm H}\sim21.5-23$, due to a decrease in the number of AGN above their effective Eddington limit, but would be weaker/disappear for the higher column densities log $N_{\rm H}\gtrsim23.5$. If we assume a typical black hole mass of $10^{8}M_{\odot}$, this means that with our boundary region, we are predicting a decrease in the fraction of log $N_{\rm H}\sim21.5-23$ objects for luminosities higher than $L_{\rm X}\approx3\times10^{43}$erg s$^{-1}$. The AGN obscuration fraction has been investigated many times, and recently with XMM observations in the COSMOS field by \cite{brusa10}. They show a decrease in the fraction of obscured sources with increasing luminosity above $L_{\rm X}\sim3\times10^{43}$erg s$^{-1}$, in agreement with previous work on this subject (e.g. \citealt{ueda03,hasinger08}) and the reasoning above.

\begin{figure}
\begin{center}
\includegraphics[width=0.8\columnwidth]{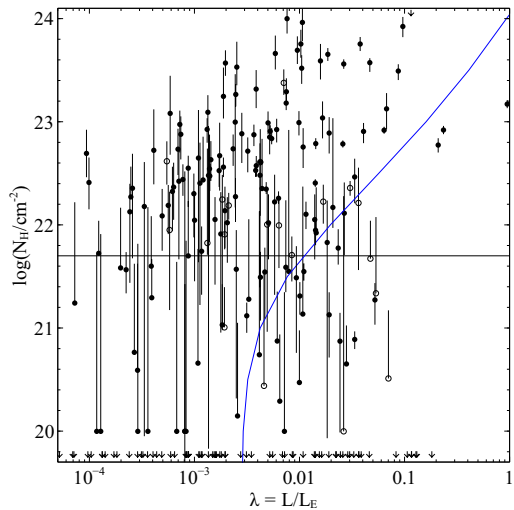}
\caption {Similar to Fig.~\ref{NhEddlowz}, but considering a redshift evolution in the black hole mass - galaxy scaling relation from \citet{merloni10}.}
\label{evol}
\end{center}
\end{figure}

Several authors have studied the evolution with redshift of galaxy scaling relations. \cite{peng06} found that the observed $M_{\rm BH}-L_{\rm R}$ relation evolves with redshift for a sample of $1.7<z<4.5$ AGN, but it is consistent with no evolution when looking at lower redshift objects ($z<1.7$). \cite{treu07} investigates this trend using observations of 20 Seyfert galaxies at $z = 0.36$, and finds an evolution in the $M_{\rm BH}-L_{\rm B}$ scaling relation. More recently, \cite{merloni10} studied a sample of type-1 AGN from the zCOSMOS survey in the redshift range $1<z<2.2$, and concluded that the objects at higher redshift tend to be brighter in the $K$-band for a given black hole mass, than the local ones.
The amount of evolution in the black hole/galaxy scaling relations is still arguable, but we aim to evaluate the effect of such evolution in our mass estimates. Considering the trend found in \cite{merloni10}, we correct our K-band luminosities by a factor $-0.73$ log(1+$z$). Our mass estimates will then be: log$($M$_{\rm BH}/$M$_{\odot})=-0.37(\pm 0.04)($M$_{\rm K} + 2.5\times0.73$ log$(1+z) + 24) + 8.29(\pm 0.08)$, which means that at our highest redshift ($z = 1$), $M_{\rm BH}$ is $\sim 1.6$ times lower than the estimate using the local scaling relation. As we can see in Fig.~\ref{evol}, these lower values for the $M_{\rm BH}$ cause the Eddington ratios to be higher, but the main conclusions of the present work remain the same. Note that the scaling found in \cite{merloni10} is valid for more luminous and higher redshift sources than our sample. Our analysis is indicative of how the results are affected by a redshift evolution, but is subject to uncertainties related with the precise form of the evolution for our type of sources.
The redshift evolution of the radiation pressure and gas interaction itself is still hard to constrain. We are limited by selection biases: low luminosity AGN are not detected at high redshifts and gas/dust can obscure luminous objects beyond our detectability limit. We hope to be able to determine how objects evolve in the $N_{\rm H}$ vs $\lambda$ plot with future X-ray surveys. The planned \emph{International X-Ray Observatory} (IXO), will be able to detect AGN outflows at high redshifts, due to its high spectral resolution at low X-ray energies ($< 2$ keV) and large effective area, by measuring the blueshift of narrow absorption lines. The sources lying close to the effective Eddington limit in our plot, and in particular, the ones clearly above that limit, are good targets for future spectroscopic studies like those.
\subsection{High Eddington ratio}
\begin{figure}
\begin{center}
\includegraphics[width=0.8\columnwidth]{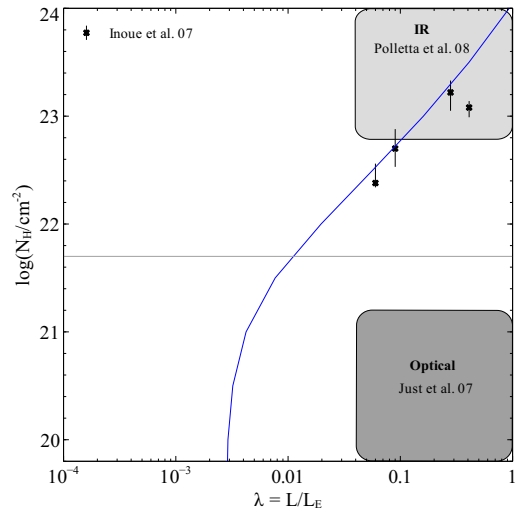}
\caption {Approximate parameter space covered by studies of luminous quasars. Light grey region represents quasars selected in the mid-IR, while the dark grey shows area covered by optical identification of luminous quasars. Crosses represent low redshift quasars from the work of \citet{inoue07}, showing signs of outflows in their spectra.}
\label{regions}
\end{center}
\end{figure}

Most of our objects lie below $\lambda\sim 0.1$ which means that the range of high Eddington ratios is not fully probed by the CDF AGN. To sample the area of $\lambda\gtrsim 0.1$ requires that we look at the regions probed by investigations of the rarer luminous quasars (Fig.~\ref{regions}). \cite{kollmeier06} and \cite{steinhardt&elvis10} show that the most luminous objects in the SDSS have $\lambda\sim0.2 - 0.3$, but do not investigate any absorption properties. \cite{greene09} look at a large sample of narrow line AGN, and find that they have high Eddington ratios ($\sim0.2$). Although these sources are obscured, there are no available values for their hydrogen column densities. \cite{just07} do study the X-ray properties of 59 of the most optically luminous quasars and find that each is consistent with no intrinsic absorption and collectively are obscured by a mean column density $<2\times10^{21}$cm$^{-2}$. For typical values of $L_{\rm X}\gtrsim10^{45}$, and imposing a maximum value of $10^{10}$M$_{\odot}$ for the black hole mass, the minimum value for $\lambda$ is $\sim0.04$. The study of optically luminous quasars covers the area of high Eddington ratios but mildly obscured objects (the dark grey region in Fig.~\ref{regions}). 
Another study of luminous quasars by \cite{shemmer08} gives upper limits for the hydrogen column density for five luminous high redshift quasars and concludes that $N_{\rm H}$ is lower than $3\times10^{21} - 4\times10^{22}$cm$^{-2}$. As in \cite{just07}, they find no significant absorption in their sources, with the data being consistent with zero intrinsic absorption. 
On the other hand, infrared-selected luminous quasars cover the high $N_{\rm H}$ range. \cite{polletta08} investigate the obscuration properties of luminous quasars selected in the mid-infrared and, using X-ray measurements, determine the column density for five of them. Their bolometric luminosity is $\sim10^{46}-10^{47}$erg s$^{-1}$, and with the same assumptions as above, the Eddington range probed is $\gtrsim 0.04$. The column densities are higher than $N_{\rm H} = 6.3\times10^{22}$cm$^{-2}$ for four of the objects, populating the light grey area plotted in Fig.~\ref{regions}, while the fifth lies in the \cite{just07} region. We also plot data from \citealt*{inoue07}, on low redshift quasars, for which $\lambda$ and $N_{\rm H}$ values are available. These objects have $L\gtrsim10^{45}$, and tend to lie close to the effective Eddington limit boundary. As predicted in the enhanced radiation pressure model, they show the presence of outflows in the UV \citep{brandtlaorwills00,laor&brandt02}. Individual X-ray spectroscopic studies of these quasars have found the presence of warm absorbers (ionised material): \cite{piconcelli04} for PG 2214+139, \cite*{brinkmann04} also for PG 2214+139 and PG 1411+442, \cite{miller06} for the broad absorption line quasar PG 1004+130 and by \cite{schartel05} and \cite{ballo08} for PG 1535+547. 
In summary, the $N_{\rm H}$ sampling of the high Eddington ratio region is not as good as for the low Eddington ratios. We have found no objects in the region bounded by $2\times10^{21}\lesssim N_{\rm H}\lesssim 4\times 10^{22}$cm$^{-2}$ and $\lambda > 0.04$ which do not show outflows, in agreement with enhanced radiation pressure arguments. There are possible selection effects on probing this region, optical measurements are affected by reddening so they might not be immediately selected as bright quasars, yet the column densities may be too low for them to be selected in the infrared. 
\subsection{Radiation pressure effect on the bulge of the galaxy}
In the scenario described in the previous sections, one can investigate the interaction between the enhanced radiation pressure and the dusty gas, and what is the long term effect on the bulge of the galaxy.
If we start by considering the Eddington luminosity for dusty gas, we have:
\begin{eqnarray*} \label{effedd}
L_{E_{d}}=\frac{4\pi Gm_{p}cM_{BH}}{\sigma_{d}}
\end{eqnarray*}
which is $\sigma_{d}/\sigma_{T}$ times lower than the Eddington luminosity for Thomson scattering. Considering a scenario where the gas is concentrated in a shell at the edge of the galaxy of radius $R$, after being swept out by a quasar at its effective Eddington limit \citep{fabianproc09}, we can balance the radiation pressure and the gravitational force:
\begin{eqnarray*} \label{balance}
\frac{L_{E_{d}}}{c}=\frac{GM_{gal}M_{gas}}{R^{2}}
\end{eqnarray*}
The mass of gas will be a fraction $f$ of the galaxy mass, and the effective Eddington ratio will now include the mass of the entire galaxy ($M_{\rm gal}\gg M_{\rm BH}$):
\begin{eqnarray*} \label{balance2}
\frac{4\pi Gm_{p}M_{gal}}{\sigma_{d}}\sim\frac{GfM^{2}_{gal}}{R^{2}}
\end{eqnarray*}
Solving this equation in order to determine $M_{gal}$, and assuming an isothermal galaxy, with $M_{gal}=2\sigma^{2}R/G$, where $\sigma$ is the velocity dispersion of the galaxy and $R$ the radius:
\begin{eqnarray} \label{mgal}
M_{gal}\sim\frac{4\pi m_{p}R^{2}}{f\sigma_{d}}\sim \frac{2\sigma^{2}R}{G}
\end{eqnarray}
\begin{eqnarray} \label{mgal2}
M_{gal}\sim\frac{f\sigma^{4}\sigma_{d}}{\pi m_{p}G^{2}}.
\end{eqnarray}
The mass of the black hole correlates with the velocity dispersion of the galaxy: $M_{\rm BH}\propto \sigma^{4}$, and it can be explained based on momentum balance arguments \citep{fabian99,fabianwilmancrawford02,king03,murray05}. Using Eq.~\ref{mgal2} and the latter correlation, we obtain a relation between the mass of the galaxy and the mass of the black hole: $M_{\rm BH}/M_{\rm gal}\propto \sigma^{-1}_{d}$. The effective cross-section ($\sigma_{d}$) is expected to decrease at higher redshifts, when the dust content in the tens of parsecs scale is lower, (see \citealt{fabian09} for details on how $\sigma_{d}/\sigma_{T}$ changes with metallicity) in agreement with studies that find an increase in the $M_{\rm BH}/M_{\rm gal}$ fraction at higher redshifts (e.g.: \citealt{merloni10}).
From Eq.~\ref{mgal}, we obtain \citep{fabianproc09}:
\begin{eqnarray*} \label{fplane}
\frac{\sigma^{2}}{R} \sim \frac{2\pi m_{p}G}{f\sigma_{d}}
\end{eqnarray*}
which indicates a relation between the velocity dispersion and the radius of the galaxy, implying a close interaction between the feedback from the AGN and the large scale properties of the galaxy. This relation is also in agreement with the dependency observed in terms of a fundamental plane \citep{hopkins07}, where if $M_{\rm BH}\propto \sigma^{4}$, then $\sigma^{2} \propto R$. 
\section{Conclusions}
We have examined the absorption column densities and Eddington fraction distributions of a large sample of AGN from the CDF-N and CDF-S fields. These objects tend to lie below their effective Eddington limit for dusty gas, with more than $95$ per cent of AGN with $N_{\rm H}>5\times 10^{21}$cm$^{-2}$, being sub-Eddington in that sense. This conclusion is in agreement with previous results, applicable to not only $z\sim(0.5-1.0)$ AGN but also to local samples. Future spectroscopic studies should find outflows at the effective Eddington limit boundary. 
Our results are consistent with the action of radiation pressure on the dusty interstellar gas within the inner 100 pc or so of the host galaxy. Much deeper studies of the rarer objects near the effective Eddington limit are required to determine the role this mechanism plays in AGN feedback. It must occur, but the effect of transient jets or fast winds could be greater. 
The lack of a significant population above the limit indicates that large amplitude changes in $\lambda$ on timescales less than the gas expulsion time ($\sim 10^{5}$yr) are rare. The overall evolution of AGN in the $\lambda$ vs $N_{\rm H}$ plane is unclear, although it is likely that an active phase is initiated when gas is plentiful and thus at high $N_{\rm H}$. If AGN exceed the effective Eddington limit, they expel much of their fuel supply (except what remains in the accretion disk) and so are likely to be short lived ($\lesssim 10^{7}$yr).

In this work, we have also suggested a possible galaxy evolution scenario in line with what is observed. The effective cross-section for dusty gas can be up to a factor $\sim1000$ higher than that for Thomson scattering, resulting in an enhanced radiation pressure capable of exceeding the gravitational force of the gas clouds around the AGN. The interaction between these two forces affects the gas content of the host galaxy, resulting in correlations between the black hole and the host galaxy, which will ultimately lead to a ``fundamental plane'' relation.

\section{Acknowledgements}
The authors thank the referee for the useful comments that improved this paper, Paul Hewett for interesting discussions, Andrea Merloni for providing details on his recent work, and Jeremy Sanders for the use of the plotting package Veusz v1.7. The authors acknowledge financial support from FCT - Funda\c c\~ao para a Ci\^encia e a Tecnologia - Portugal (SR), The Royal Society (ACF and DMA), a Philip Leverhulme Prize (DMA), CXC grant SP8-9003A and NASA ADP grant NNX10AC99G (WNB, BL, RVV and YQX). 
\bibliographystyle{mnras}
\bibliography{AGN}
\end{document}